%% file: template.tex
\documentclass{vgtc}                          

\ifpdf
  \pdfoutput=1\relax                   
  \usepackage{graphicx}                
  \DeclareGraphicsExtensions{.pdf,.png,.jpg,.jpeg} 
\else
  \usepackage{graphicx}                
  \DeclareGraphicsExtensions{.eps}     
\fi%

\graphicspath{{figures/}{pictures/}{images/}{./}} 

\usepackage{microtype}                 
\PassOptionsToPackage{warn}{textcomp}  
\usepackage{textcomp}                  
\usepackage{mathptmx}                  
\usepackage{times}                     
\usepackage{cite}                      
\usepackage{tabu}                      
\usepackage{booktabs}                  
\usepackage{hyperref}

\onlineid{0}

\vgtccategory{Research}

\vgtcinsertpkg

\title{Combinational Nonuniform Timeslicing of Dynamic Networks}

\author{Seokweon Jung\thanks{e-mail: swjung@hcil.snu.ac.kr}\\ %
    \scriptsize Seoul National University %
\and DongHwa Shin\thanks{e-mail: dhshin@kw.ac.kr}\\ %
    \scriptsize Kwangwoon University %
\and Hyeon Jeon\thanks{e-mail: hj@hcil.snu.ac.kr}\\ %
    \scriptsize Seoul National University %
\and Jinwook Seo\thanks{e-mail: jseo@snu.ac.kr, corresponding author}\\ %
    \scriptsize Seoul National University%
}
\vspace{-3cm}
\teaser{
  \centering
  \includegraphics[width=\linewidth]{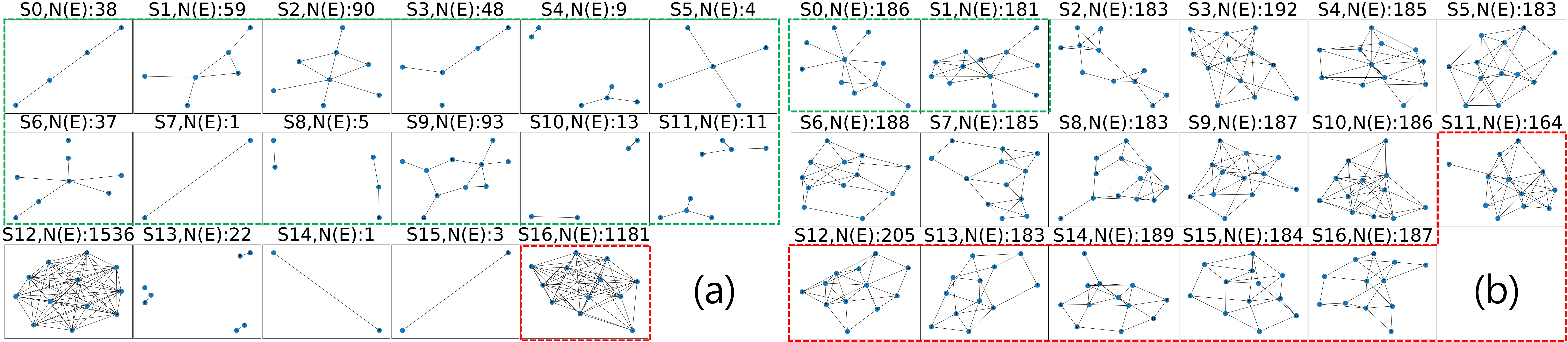}
  \vspace{-0.7cm}
  \caption{Result of two different nonuniform time slicing of Rugby dataset: (a) Jaccard similarity-based slicing (JS), and (B) Visual complexity-based slicing (VS). A dynamic network is divided into 17 snapshots with their index and number of included events marked at the top of each snapshot. JS aggregates similar adjacent temporal segments into a single snapshot, resulting in a dense snapshot while VS splits it into six snapshots (red boxes). JS also creates twelve sparse and noisy snapshots when the network changes abruptly, while VS merges them into two stable snapshots (green boxes).}
  \label{fig:teaser}
}

\abstract{
Dynamic networks represent the complex and evolving interrelationships between real-world entities. Given the scale and variability of these networks, finding an optimal slicing interval is essential for meaningful analysis. Nonuniform timeslicing, which adapts to density changes within the network, is drawing attention as a solution to this problem. In this research, we categorized existing algorithms into two domains---data mining and visualization---according to their approach to the problem. Data mining approach focuses on capturing temporal patterns of dynamic networks, while visualization approach emphasizes lessening the burden of analysis. We then introduce a novel nonuniform timeslicing method that synthesizes the strengths of both approaches, demonstrating its efficacy with a real-world data. The findings suggest that combining the two approaches offers the potential for more effective network analysis.
} 

\CCScatlist{
  \CCScatTwelve{Human-centered computing}{Visualization}{Visualization techniques}{Graph drawings};
}

\begin{document}


\input{sections/01_introduction}
\input{sections/02_method}

\input{sections/03_evaluation}

\input{sections/04_conclusion}

\acknowledgments{
This work was supported by the National Research Foundation of Korea (NRF) grant funded by the Korea government (MSIT) (No. 2023R1A2C200520911).
}

\bibliographystyle{abbrv-doi}

\bibliography{template}
\end{document}

%% file: sections/01_introduction.tex
\firstsection{Introduction}

\maketitle

The various entities existing in real life and the relationships between them can be represented in the form of a dynamic network that changes over time.
Because these dynamic networks contain various information, such as relationships and communities within the network, many companies or academia are analyzing them for various purposes.
Because of the large scale and complexity of dynamic networks, it has been a primary goal to find out the optimal slicing interval for effective analysis of dynamic networks. However, the uniform interval falls short of accommodating the variable density and other real-world dynamics intrinsic to these networks~\cite{Orman2021-cv}. 
This discrepancy has led to the proposition of nonuniform slicing—an adaptive approach that modifies slice intervals in response to network changes. 
Summarizing existing literature, research on nonuniform dynamic network slicing bifurcates into two main branches: data mining-based and visualization-based.

\textbf{Data mining} approach first divides a dynamic network into minimal segments whose interval is \(\Delta t\). Then, with various network analysis measures such as Jaccard similarity~\cite{Orman2021-cv, Colak2021-ys}, edge probability~\cite{Wendt2023-bx}, and community structure~\cite{Luo2023-na}, aggregates adjacent segments whose similarities are above the threshold.
It is akin to change point detection, particularly because it segments snapshots at points where similarity decreases. This approach carries the advantage that each snapshot inherently consists of segments with high internal similarity, thereby clearly describing common patterns~(\autoref{fig:teaser}(a), red box). Consequently, it also allows for a distinct visualization of network changes, providing a clear comparison between before-and-after of transition. However, when a dynamic network changes abruptly, the slicing result can be noisy and unstable, including only a few edges per snapshot~(\autoref{fig:teaser}(a), green box).

\textbf{Visualization} approach emphasizes reducing cognitive load during visual analysis. There exist various visualization methods, such as space-time mapping~\cite{Ponciano2021-nh}, animation~\cite{Crnovrsanin2021-co}, multivariate network visualization~\cite{arleo2022event}, and corresponding nonuniform slicing algorithms to present temporal changes of dynamic networks. Small multiples is one of the widely accepted visualization methods.
Researchers tried to minimize the visual complexity among multiple timeslicies, which causes the cognitive load, to achieve the goal. For instance, Wang et al.~\cite{Wang2019-qf} brought the notion of histogram equalization to prevent too sparse or dense snapshots~(\autoref{fig:teaser}(b))

Despite the advancements in both domains, a gap persists between the data mining focus on helping effective pattern detection through slicing and the visualization emphasis on lessening the burden of analysis. As there exists a need to fulfill both goals, both must be treated with importance~\cite{jung2023monetexplorer}.

In this paper, we present a novel nonuniform timeslicing method of dynamic networks that integrates the strengths of two distinct methodologies, thereby mitigating their respective weaknesses. Based on our analysis and observations about existing algorithms, we propose a simple and effective method of combining algorithms from each domain. With a real data, we present that the suggested algorithm can provide a better network analysis experience.

%% file: sections/02_method.tex
\section{Combinational Nonuniform Timeslicing}
We propose a nonuniform timeslicing algorithm combining data mining-based and visualization-based approaches.
The inspiration came from the observed characteristics of each approach. Firstly, the data mining practice of segmenting and then aggregating neighboring segments based on similarity~\cite{Orman2021-cv, Colak2021-ys, Luo2023-na}.
Secondly, the phenomenon that visualization approaches tend to yield a sequence of similar types.
For instance, snapshots in the red dotted box of \autoref{fig:teaser}(b) share a similar network structure.
Our approach involves:

\vspace{-0.1cm}
\begin{itemize}
    \item \textbf{Step 1:} Using a visual complexity method to create fine-grained temporal segments from a dynamic network.
    \vspace{-0.1cm}
    \item \textbf{Step 2:} Employing a data mining method to aggregate these segments into nonuniform snapshots.
\end{itemize}
\vspace{-0.1cm}

We implemented our algorithm with the nonuniform time slicing methods suggested by \c{C}olak et al.~\cite{Colak2021-ys}, and Wang et al.~\cite{Wang2019-qf}.
Wang's method defines visual complexity as the number of edges present in a snapshot and uses histogram equalization to distribute these across snapshots evenly. \c{C}olak's method performs aggregation between adjacent segments to form snapshots based on the Jaccard similarity of link sets.
We will refer to the former as Visual complexity-based slicing (VS) and the latter as Jaccard similarity-based slicing (JS).

%% file: sections/03_evaluation.tex
\begin{figure}[t]
  \centering
  \includegraphics[width=\linewidth]{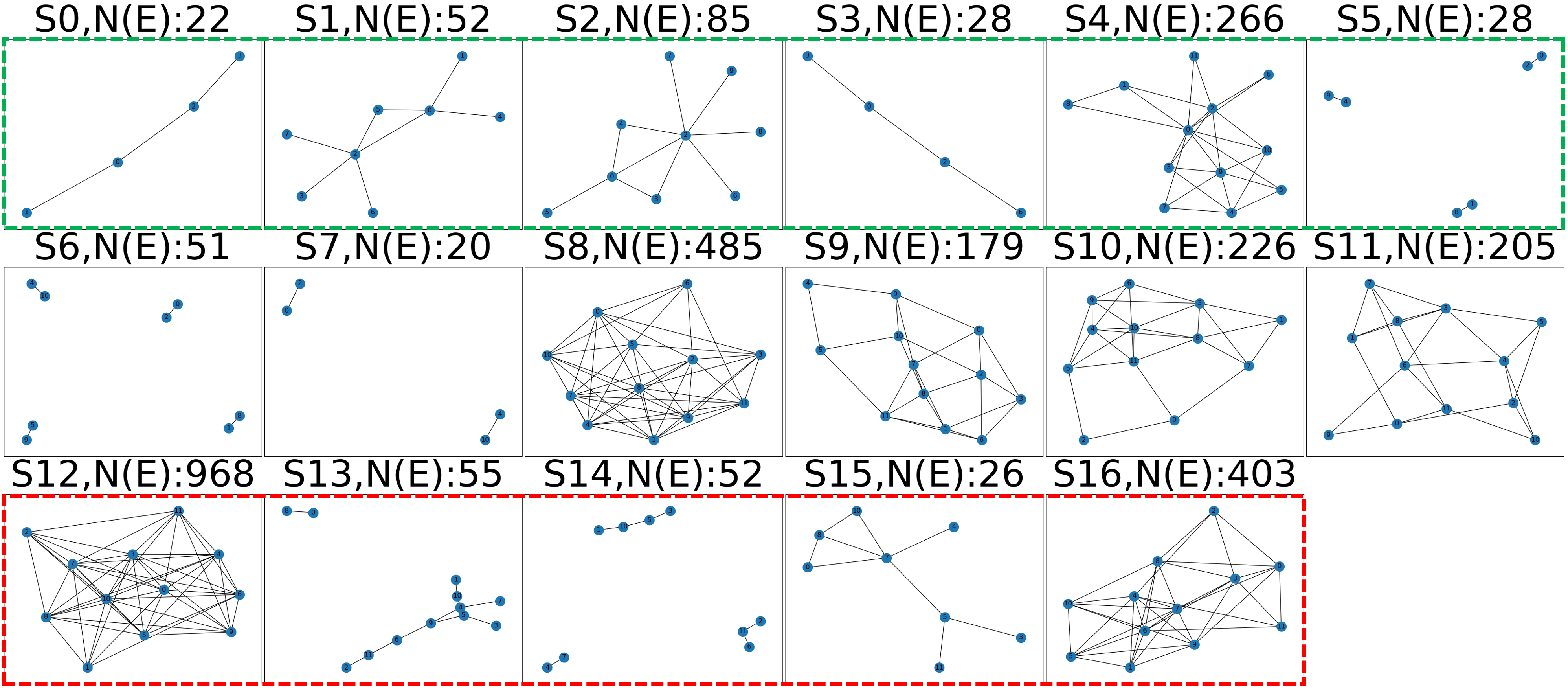}
\vspace{-0.7cm}
  \caption{Result of the combinational nonuniform timeslicing utilizing both Jaccard similarity-based and visual complexity-based slicing method. The noise at the beginning reduced as the number of events in each snapshot increased while showing significant changes over time (green box). Snapshots at the end now clearly visualize the final evolution of the network, revealing a hidden pattern (red box).}
  \label{fig:combined}
  \vspace{-0.2cm}
\end{figure}

\section{Evaluation}
We used the Rugby Dataset~\cite{simonetto2018event} to demonstrate the effectiveness of our technique. The dataset contains over 3,000 tweets between the 12 Rugby teams in the Guinness Pro12 competition. Each tweet consists of involved teams and a timestamp. As VS requires the number of snapshots as a direct input, we first performed JS and then adjusted VS to match the resulting number of snapshots. 
JS takes the size of the temporal segment, \(\Delta t\), as a parameter. After testing various  \(\Delta t\), we determined that a week is the proper  \(\Delta t\), producing 17 snapshots that allow for discernible patterns over time. Consequently, VS was calibrated to generate an equivalent set of 17 snapshots.
Each snapshot is numbered from 0 to 16.

Consistent with their respective designs, JS highlighted snapshots with dense, specific patterns where tweets were concentrated~(\autoref{fig:teaser}(a), red box), while in contrast, it revealed noisy patterns in sparser regions~(\autoref{fig:teaser}(a), green box). On the other hand, VS demonstrated a more even distribution of tweets across all snapshots~(\autoref{fig:teaser}(b)).

Then, we implemented a combinational nonuniform timeslicing with JS and VS. 
We first generated 118 preliminary snapshots using VS, then aggregated them using JS to produce 17 final snapshots. The generated snapshots are shown in ~\autoref{fig:combined}. Upon comparison with the baseline methods, we observed that the area marked with a green box was sparser than what VS produced yet denser than JS, allowing for the observation of changing patterns. In the area marked with a red box, our approach shows a temporal change network structure across five snapshots, while all details were abstracted into a single snapshot with JS, making it difficult to discern specific patterns. Although fewer in number compared to VS, six to five, this combined method has the advantage of displaying more distinct and detailed patterns.

%% file: sections/04_conclusion.tex
\section{Conclusion and Future Work}

In this paper, we have explored two distinct approaches to addressing the issue of nonuniform time slicing in dynamic networks.
We classified algorithms into two classes and implemented one method from each approach for a comparative analysis with a real data to understand their advantages and disadvantages during the visual analysis process.
Our findings reveal that the benefits of each methodology depend significantly on the level of detail researchers wish to observe within the network. Additionally, we discovered that by combining the opposing approaches, it is possible to mitigate some of the issues inherent to each individual method.

We further aim to develop an advanced nonuniform timeslicing technique based on our findings. This includes applying various nonuniform timeslicing methods from the data mining and visualization field. Since the values that each method values are different, the results can be significantly different. Furthermore, we plan to expand our research interest to not only how to slice the dynamic network but also how to evaluate the result. The absence of a concrete evaluation metric of "good snapshot" creates a gap in the evaluation of slicing results. The establishment of concrete evaluation metrics will extend our understanding of how to best represent dynamic networks in a way that facilitates deeper insights and understanding.

%% file: template.bbl
\begin{thebibliography}{10}

\bibitem{arleo2022event}
A.~Arleo, S.~Miksch, and D.~Archambault.
\newblock Event-based dynamic graph drawing without the agonizing pain.
\newblock In {\em Computer Graphics Forum}, vol.~41, pp. 226--244. Wiley Online Library, 2022.

\bibitem{Colak2021-ys}
S.~{\c C}olak and G.~K. Orman.
\newblock Aggregating time windows for dynamic network extraction.
\newblock In {\em 2021 International Conference on {INnovations} in Intelligent {SysTems} and Applications ({INISTA})}, pp. 1--6, Aug. 2021.

\bibitem{Crnovrsanin2021-co}
T.~Crnovrsanin, {Shilpika}, S.~Chandrasegaran, and K.-L. Ma.
\newblock Staged animation strategies for online dynamic networks.
\newblock {\em IEEE Trans. Vis. Comput. Graph.}, 27(2):539--549, Feb. 2021.

\bibitem{jung2023monetexplorer}
S.~Jung, D.~Shin, H.~Jeon, K.~Choe, and J.~Seo.
\newblock Monetexplorer: A visual analytics system for analyzing dynamic networks with temporal network motifs.
\newblock {\em IEEE Transactions on Visualization and Computer Graphics}, 2023.

\bibitem{Luo2023-na}
X.~Luo, T.~Wang, G.~Xin, Y.~Lu, K.~Yan, and Y.~Liu.
\newblock {Classifier-Based} nonuniform time slicing method for local community evolution analysis.
\newblock {\em Big Data Research}, p. 100408, Sept. 2023.

\bibitem{Orman2021-cv}
G.~K. Orman, N.~T{\"u}re, S.~Balcisoy, and H.~A. Boz.
\newblock Finding proper time intervals for dynamic network extraction.
\newblock {\em J. Stat. Mech: Theory Exp.}, 2021(3):0--0, Mar. 2021.

\bibitem{Ponciano2021-nh}
J.~R. Ponciano, C.~D.~G. Linhares, E.~R. Faria, and B.~A.~N. Traven{\c c}olo.
\newblock An online and nonuniform timeslicing method for network visualisation.
\newblock {\em Comput. Graph.}, 97:170--182, June 2021.

\bibitem{simonetto2018event}
P.~Simonetto, D.~Archambault, and S.~Kobourov.
\newblock Event-based dynamic graph visualisation.
\newblock {\em IEEE Transactions on Visualization and Computer Graphics}, 26(7):2373--2386, 2018.

\bibitem{Wang2019-qf}
Y.~Wang, D.~Archambault, H.~Haleem, T.~Moeller, Y.~Wu, and H.~Qu.
\newblock Nonuniform timeslicing of dynamic graphs based on visual complexity.
\newblock In {\em 2019 {IEEE} Visualization Conference ({VIS})}, pp. 1--5, Oct. 2019.

\bibitem{Wendt2023-bx}
J.~D. Wendt, R.~Field, C.~Phillips, A.~Prasadan, T.~Wilson, S.~Soundarajan, and S.~Bhowmick.
\newblock Partitioning communication streams into graph snapshots.
\newblock {\em IEEE Transactions on Network Science and Engineering}, 10(2):809--826, 2023.

\end{thebibliography}
